%% file: main.tex
\documentclass[aps,notitlepage,superscriptaddress,floatfix,pra,twocolumn]{revtex4-1}

\usepackage{lipsum}
\usepackage{graphicx}
\usepackage{amsmath,amssymb,mathrsfs}
\usepackage{braket}
\usepackage{mathtools}
\usepackage{hyperref}
\hypersetup{
    colorlinks=true,
    citecolor=blue!60!black,
    linkcolor=black,
    urlcolor=blue!60!black
}
\usepackage{multirow}
\usepackage{color}
\usepackage[normalem]{ulem}
\usepackage{amsfonts}
\usepackage{float}
\usepackage{pdfpages} 
\makeatletter
\usepackage{subfigure}
\usepackage{soul, xcolor}
\usepackage{dsfont}

\DeclareMathOperator{\Tr}{\mbox{Tr}}

\newcommand{\dd}{d}

\setstcolor{red}

\def\beq{\begin{equation}}
\def\eeq{\end{equation}}
\def\bea{\begin{eqnarray}}
\def\eea{\end{eqnarray}}

\usepackage{pdfpages} 
\makeatletter
\AtBeginDocument{\let\LS@rot\@undefined}
\makeatother

\begin{document}

\title{Typicality of entanglement growth and transport fluctuations in\\ quantum symmetric exclusion processes}

\author{Ewan McCulloch}
\affiliation{Laboratoire de Physique de l'École Normale Supérieure, CNRS, ENS \& Université PSL; 24 rue Lhomond, 75005 Paris, France}

\begin{abstract}
We use macroscopic fluctuation theory (MFT) to establish the typicality of charge-transfer statistics and entanglement dynamics in the quantum symmetric exclusion process, a model of stochastically hopping free fermions with diffusive transport and long-range quantum coherence. We show that the leading long-time charge-transfer full counting statistics of individual noise realizations coincide almost surely with those of the classical symmetric exclusion process (SSEP), in equilibrium and following quenches from arbitrary occupation-diagonal product states. Our results apply to transient fluctuations in the thermodynamic limit, extending previous work that has primarily focused on stationary regimes. Our approach compares the noise averages of the charge-transfer generating function and its inverse. The former reduces to SSEP, while the latter maps to an auxiliary Brownian energy process whose MFT saddle-point problem is related to that of SSEP, yielding matching large-deviation exponents. Through free-fermion identities relating entanglement to subsystem-number fluctuations, this establishes the self-averaging of the covariance-matrix eigenvalue distribution and of Rényi and von Neumann entropies. We obtain asymptotically exact predictions for entanglement growth from domain-wall and Néel initial states, in excellent agreement with numerical simulations.
\end{abstract}

\maketitle

\textbf{Introduction}---Hydrodynamic large deviations are emerging as an organizing principle for intrinsically quantum phenomena in many-body systems. In chaotic systems, recent work has shown that local quantum correlations are controlled by ``voids''---rare density fluctuations in which chaotic dynamics is slowed---with macroscopic fluctuation theory (MFT)~\cite{Bertini2002,PhysRevLett.94.030601,Bertini_2015,Derrida_2025} enabling quantitative predictions for their decay~\cite{first_coherence_paper,previous,McCullochKPZ2026}. Similar rare density fluctuations are also responsible for the anomalously slow diffusive growth of higher Rényi entropies observed in chaotic systems~\cite{PhysRevLett.122.250602,PhysRevResearch.2.033020,Huang_2020,Rakovszky2021}. Yet, despite recent progress in interacting Brownian Hamiltonian systems~\cite{Lastres2026}, a systematic theory of entanglement based on hydrodynamic large deviations is still missing.

Remarkably, hydrodynamic fluctuations have also been connected to quantum observables in integrable systems, but by a completely different mechanism. For free-fermion integrable systems, both entanglement and dynamical spin-coherence correlations depend on the full distribution of charge fluctuations across a cut, rather than merely its first few cumulants~\cite{SciPostPhysCore.7.1.005,DelVecchio2022,PhysRevLett.102.100502}. More precisely, they probe the corresponding cumulant generating function (CGF) at finite counting fields. These quantum observables therefore probe the rare fluctuations in transport directly accessible to hydrodynamic large-deviation theories—MFT and its recently developed ballistic counterpart, ballistic fluctuation theory~\cite{doyon2020fluctuations,doyon2023ballistic}.

In integrable systems, transport fluctuations are inherited from the initial state and propagated deterministically by generalized hydrodynamics and its quantum extension~\cite{PhysRevX.6.041065,PhysRevLett.117.207201,Myers2018,PhysRevLett.122.090601,SciPostPhys.10.5.116,Malvania2021,PhysRevX.15.010501,PhysRevLett.124.140603,Ruggiero_2022,PhysRevA.108.013324,Scopa_2022}. In diffusive systems, by contrast, the dynamics is effectively stochastic---the noisy diffusion equation
is expected to describe both classical and quantum
chaotic systems~\cite{PhysRevA.89.053608,Bohrdt_2017,Gu2017,Rakovszky2018,Khemani_2018,PhysRevB.103.115148,PhysRevB.105.075131,PhysRevB.95.155131,SciPostPhys.2.3.018,McCulloch2023,wienand2023emergence,Singh2025,Christopoulos2026}---so large deviations in transport involve a different optimization, over density and current histories as well as initial fluctuations~\cite{Derrida_2009,DerridaGerschenfeld2009MFT,Krapivsky_2012,DerridaII2019,Dandekar_2022,Mallick_2022,PhysRevLett.133.117102,Berlioz2025,10.21468/SciPostPhys.17.2.033,Sharma2026,Smith2024,Grabsch2026,Berlioz_2024}.


The quantum simple symmetric exclusion process (QSSEP)~\cite{BauerBernardJin2019,Bernard_2019,QSSEP_Bernard_2021} provides a particularly tractable setting in which to use MFT to make quantitative predictions for entanglement dynamics in a diffusive quantum system. This model of stochastically hopping free fermions exhibits diffusive transport while supporting long-range quantum coherence in boundary-driven steady states (which is generically short-ranged in chaotic systems). Its noise-averaged dynamics reduces to the classical symmetric simple exclusion process (SSEP), and the corresponding fluctuating hydrodynamics is expected to hold for typical realizations. Previous results establishing classical fluctuation
statistics in individual realizations have
focused on the long-time stationary regime of
boundary-driven QSSEP~
\cite{Bernard2025,Costa2026,Albert2026}.
Entanglement in QSSEP has likewise been studied primarily in late-time stationary states~\cite{BernardPiroli2021,SciPostPhys.15.4.175}.
More recently, operator entanglement growth was calculated by assuming the self-averaging of charge-transfer statistics~\cite{tonetto2026}.
By contrast, the typicality of transient charge-transfer statistics, both in equilibrium and following quantum quenches, as well as of entanglement dynamics remains an open problem~\footnote{Refs.~\cite{BernardPiroliScopa2026,Bayona2026}
and~\cite{Alba2026} study entanglement dynamics in related noisy free-fermion models, with all-to-all hopping and spatially homogeneous Brownian hopping, respectively. In the former, transport is nonlocal and non-diffusive. In the latter, quasiparticle motion is governed by the statistics of a single random walk rather than the emergent fluctuating diffusion described by SSEP hydrodynamics.}.

\begin{figure}[t]
    \centering
    \includegraphics[width=0.85\linewidth]{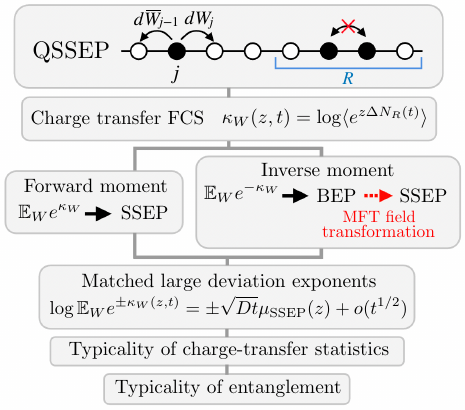}
    \vspace{-2mm}
\caption{\textbf{Schematic of the typicality derivation}. The forward and
inverse moments of the QSSEP charge-transfer generating function map to classical SSEP and a Brownian energy process (BEP)~\cite{GiardinaEtAl2009}, respectively.
A field transformation relates their MFT saddle problems, yielding matching large-deviation functions and establishing typicality of charge-transfer statistics,
and by Ref.~\cite{PhysRevLett.102.100502}, entanglement dynamics.}
\vspace{-3mm}
\label{fig:typicalitySchematic}
\end{figure}

We introduce a new strategy for establishing the
typicality of transient charge-transfer statistics,
illustrated in Fig.~\ref{fig:typicalitySchematic}. We compare the noise averages of the charge-transfer generating function and its inverse: the former reduces to SSEP, while the latter maps exactly to a Brownian energy process (BEP)~\cite{GiardinaEtAl2009}. A transformation of the hydrodynamic fields relates their MFT saddle problems, yielding matching large-deviation exponents. We thereby establish that the leading charge-transfer full counting statistics (FCS) coincides almost surely with that of classical SSEP, both in equilibrium and following quenches, for arbitrary occupation-diagonal product-state preparations. Exploiting the relation between free-fermion entanglement and FCS~\cite{PhysRevLett.102.100502,SongEtAl2011}, we obtain asymptotically exact predictions for the $\sqrt{t}$ growth of Rényi and von Neumann entropies in individual noise realizations.

\textbf{Entanglement from charge statistics}---A number-conserving fermionic Gaussian state has the form $\rho\propto e^{-{\cal H}}$, where ${\cal H}=\sum_{i,j}J_{ij}c_i^\dagger c_j$, with $J$ Hermitian. Such a state is fully characterized by its single-particle covariance matrix, $C_{ij}\equiv \Tr(\rho\,c_j^\dagger c_i)$. Free-fermion dynamics preserves Gaussianity: denoting the many-body evolution for a noise realization $W$ by $\mathcal U_W(t)$, its action on fermion operators $c_i(t)={\sum}_j[U_W(t)]_{ij}c_j$ defines the single-particle propagator $U_W(t)$. The evolved Gaussian state $\rho_W(t)$ is therefore characterized by $C_W(t)=U_W(t)C_0U_W(t)^\dagger$.

The reduced density matrix $\rho_{R,W}(t)=\Tr_{\bar R}\rho_W(t)$ of a subsystem $R$ is itself Gaussian, with covariance $C_{R,W}(t)$ given by the restriction of $C_W(t)$ to $R$. In the basis that diagonalizes $C_{R,W}(t)$, $\rho_{R,W}(t)$ factorizes into independent fermionic modes with occupations $\nu_a$. Thus,
\begin{equation}
\begin{aligned}
\Tr[\rho_{R,W}(t)^m]
&=\det\nolimits_R[(1-C_{R,W})^m+C_{R,W}^m],\\
\chi_W(z,t)
&\equiv\log\Tr[\rho_{R,W}(t)e^{zN_R}]\\
&=\log\det\nolimits_R[1+(e^z-1)C_{R,W}].
\end{aligned}
\label{eq:gaussianDeterminants}
\end{equation}
For integer $m\ge 2$, the Rényi entropies are related to subsystem-number statistics by~\cite{PhysRevLett.102.100502,SongEtAl2011},
\begin{equation}
\Delta S_{m,W}=\frac{1}{1-m}\sum_{k=0}^{m-1}
\Delta\chi_W(i\theta_{m,k},t),
\label{eq:entropyFCS}
\end{equation}
where $\theta_{m,k}=(2k+1-m)\pi/m$, $\Delta\chi_W(z,t)\equiv\chi_W(z,t)-\chi_W(z,0)$, and $\Delta S_{m,W}\equiv S_{m,W}(t)-S_{m,W}(0)$. The von Neumann entropy has an analogous representation involving only real counting fields,
\begin{equation}
\Delta S_{1,W}=\frac14\int_0^\infty
\frac{\dd u}{\sinh^2(u/2)}
[\Delta\chi_W(u,t)+\Delta\chi_W(-u,t)].
\label{eq:vonNeumannFCS}
\end{equation}
(This identity follows from the same Gaussian spectrum; a derivation is given in the End Matter.)

We consider diagonal product states ${\rho_0=\bigotimes_j\rho_j}$, with $\rho_j=(1-\varrho_j)|0_j\rangle\langle0_j| +\varrho_j|1_j\rangle\langle1_j|$ and arbitrary $0\le\varrho_j\le1$.
To connect entropy dynamics to classical transport, we introduce the charge transfer generating function~\cite{Levitov1993,Levitov1996ElectronCS,RevModPhys.81.1665}
\begin{equation}
\begin{aligned}
Z_W(z,t;\rho_0)&\equiv \Tr\!\left[e^{zN_R}\mathcal U_W(t)
e^{-zN_R}\rho_0\,\mathcal U_W(t)^\dagger\right],\\
\kappa_W(z,t;\rho_0)&\equiv \log Z_W(z,t;\rho_0).
\end{aligned}
\label{eq:transferFCS}
\end{equation}
The initial counting factor can be absorbed into a tilted product state, $\rho_0(z)=e^{zN_R-\chi_W(z,0)}\rho_0$, whose site occupation probabilities are
\begin{equation}
\varrho_j(z)=
\frac{\varrho_j e^{z\mathbf 1_{\{j\in R\}}}}
{1-\varrho_j+\varrho_j e^{z\mathbf 1_{\{j\in R\}}}}.
\label{eq:tiltedinitial}
\end{equation}
The change in the subsystem-number CGF is therefore exactly the charge-transfer CGF evaluated in this tilted initial state:
\begin{equation}
\Delta\chi_W(z,t)=\kappa_W(z,t;\rho_0(z)).
\label{eq:entropytransfer}
\end{equation}
For pure occupation-product states, $\varrho_j\in\{0,1\}$, this tilt is trivial and the subsystem entropies measure bipartite entanglement.

\textbf{QSSEP and self-averaging}---The quantum symmetric simple exclusion process describes fermions hopping under independent complex Brownian noise on each bond, with stochastic Hamiltonian increment
\begin{equation}
dH_W=\sqrt D\, {\sum}_j
\left(c_{j+1}^\dagger c_j\,dW_j
+c_j^\dagger c_{j+1}\,d\overline W_j\right),
\label{eq:QSSEP}
\end{equation}
where $\mathbb E_W[dW_j d\overline W_k] =\delta_{jk}dt$ and $\mathbb E_W[dW_j dW_k]=0$.

Our central result is that the leading charge-transfer statistics of a typical QSSEP realization are governed by those of the classical SSEP. Specifically, for any occupation-diagonal product initial state $\rho'_0$, every real counting field $z$, and $1\ll \sqrt{Dt}\ll L$, we show that
\begin{equation}
 \frac{\kappa_W(z,t;\rho'_0)}{\sqrt{Dt}}
 \longrightarrow
 \mu_{\rm SSEP}(z;\rho'_0), \quad \textrm{almost surely,}\
 \label{eq:typicalgoal}
\end{equation}
where $\mu_{\rm SSEP}$ is the scaled SSEP charge-transfer CGF~\cite{Derrida_2009,Mallick_2022}, for initial site occupations specified by $\rho'_0$.

Crucially, this typicality extends to quantum entanglement and subsystem entropy dynamics. Through Eq.~\eqref{eq:entropytransfer}, the change in the subsystem-number CGF is determined by the same SSEP current statistics, evaluated in the tilted initial state $\rho_0(z)$.  Together with Eqs.~\eqref{eq:entropyFCS} and \eqref{eq:vonNeumannFCS}, this determines the leading R\'enyi and von Neumann entropy growth~\footnote{ For the von Neumann entropy, exchanging the long-time limit with the counting-field integral requires additional care. In the Supplemental Material, we justify this using a finite-time diffusion bound that controls the contribution from arbitrarily large counting fields.}. The analytic continuation is established in the End Matter for pure initial states and extended to general occupation-diagonal product states in the Supplemental Material.

\textbf{Exact entanglement predictions}---We take
$R$ to be the right half-line and consider two pure
occupation-product initial states: a domain wall, with sites to the left occupied and those to the right empty, and a N\'eel state, with alternating occupations. In both cases the initial particle number in $R$ is fixed, so that $\Delta\chi_W=\kappa_W$. For the domain wall, the SSEP current CGF is known exactly \cite{Derrida_2009,Mallick_2022}. The N\'eel state corresponds to fixed, spatially uniform initial density $n_0=1/2$. At half filling, the SSEP current CGF for this quenched initial condition is related to that for equilibrium (``annealed'') initial conditions by a factor $1/\sqrt{2}$ \cite{DerridaGerschenfeld2009MFT}. The resulting scaled CGFs are
\begin{align}
\mu_{\mathrm{DW}}(z)
&=\int_{\mathbb R}\frac{dk}{\pi}
\log[1+(e^z-1)e^{-k^2}],
\notag\\
\mu_{\textrm{N\'eel}}(z)
&=\int_{\mathbb R}\frac{dk}{\sqrt{2}\pi}
\log[1+\sinh^2(z/2)e^{-k^2}].
\label{eq:exactSSEPCGFs}
\end{align}
Equations~\eqref{eq:entropyFCS}-\eqref{eq:vonNeumannFCS} thus give $S_{m,W}^R(t)=s_m\sqrt{Dt}+o(\sqrt{Dt})$, with $(s_1,s_2,s_3)_{\mathrm{DW}}\simeq(0.5238,0.4141,0.3729)$~\cite{tonetto2026} and  $(s_1,s_2,s_3)_{\textrm{N\'eel}} \simeq(0.6363,0.4986,0.4479)$ (see Supplemental Material for details).

The SSEP charge-transfer statistics also determine
the change in the subsystem covariance spectrum. Let $\nu_a(t)$ denote the
eigenvalues of the subsystem covariance matrix.
We characterize their dynamics through the
change in the cumulative eigenvalue count, rescaled by $\sqrt{Dt}$,
\begin{equation}
\mathcal N_W(\nu,t)=\frac{1}{\sqrt{Dt}}
\sum_a\left[
\mathbf 1_{\{\nu_a(t)>\nu\}}
-\mathbf 1_{\{\nu_a(0)>\nu\}}
\right],
\label{eq:cumulativeSpectrum}
\end{equation}
where $\mathbf 1_{\{A\}}$ denotes the indicator
function, equal to $1$ when the condition $A$
holds and $0$ otherwise.

\begin{figure}
    \centering
    \includegraphics[width=0.9\linewidth]{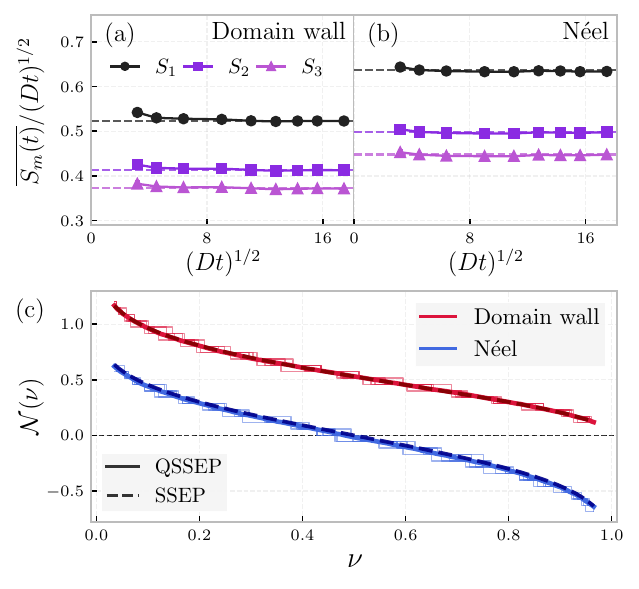}
    \vspace{-4mm}
    \caption{
    Exact entanglement predictions for (a) domain-wall and
    (b) N\'eel initial states. (a)--(b) show the noise-averaged
    R\'enyi and von Neumann entropies, rescaled by $\sqrt{Dt}$;
    dashed lines are the analytical SSEP predictions.
    (c) Rescaled change in cumulative eigenvalue count (Eq.~\eqref{eq:cumulativeSpectrum}) at
    $t=300$. Solid (dashed) curves show numerical averages (analytical predictions), with
    five realizations shown in faint lines.
    Simulations use $L=100$, $N=200$ samples and timestep $\delta t=0.05$.
    }
    \vspace{-2mm}
    \label{fig:exactPredictions}
\end{figure}

For the domain-wall and N\'eel initial states,
we obtain the limiting distributions
$\mathcal N_{\mathrm{DW}}(\nu)
=\frac{2}{\pi}\sqrt{-\log\nu}$ and
\begin{align}
\mathcal N_{\textrm{N\'eel}}(\nu)
=\frac{\sqrt{2}}{\pi}
\operatorname{sgn}\!\left(\frac12-\nu\right)
\sqrt{-\log[4\nu(1-\nu)]},
\label{eq:exactSpectra}
\end{align}
valid for $0<\nu<1$. (See End Matter for details.)
Figure~\ref{fig:exactPredictions} compares these predictions with numerical simulations, showing excellent agreement for both entanglement growth and the covariance spectrum.

\textbf{Establishing typicality}---To show typicality analytically we compute the large-time asymptotics of both the
transferred-charge generating function and its inverse,
\begin{equation}
\mathbb E_W Z_W^{\pm1}(z,t;\rho_0')
=\exp\left[\pm\sqrt{Dt}\,\mu_{\rm SSEP}(z;\rho'_0)
+o(t^{1/2})\right].
\label{eq:momentsgoal}
\end{equation}
By Markov's inequality, the matching large-deviation exponents in Eq.~\eqref{eq:momentsgoal} immediately imply convergence in probability. We strengthen this to the almost-sure result in Eq.~\eqref{eq:typicalgoal} by exploiting the locality of charge transfer: only hopping across the cut changes $Z_W$, allowing us to show that the rescaled CGF remains arbitrarily close to the SSEP prediction at all sufficiently late times, for almost every noise realization (see Supplemental Material).

The forward average follows from the well-known
equivalence between noise-averaged QSSEP dynamics
and classical SSEP for occupation-diagonal states~\cite{Bernard_2019}.
Since the initial counting factor $e^{-zN_R}$
preserves diagonality, this correspondence gives
the exact identity
\begin{equation}
\mathbb E_W Z_W(z,t;\rho'_0)
=\langle e^{z[N_R(t)-N_R(0)]}\rangle_{{\rm SSEP},\rho'_0},
\label{eq:forwardSSEP}
\end{equation}
where the SSEP has particles exchanging with
neighbouring holes at rate $D$, with independent
initial occupations of probabilities $\varrho'_j$
specified by $\rho'_0$.

In general,
$\mathbb E_W Z_W^{-1}\neq(\mathbb E_W Z_W)^{-1}$,
so the inverse average requires a separate calculation.
We show that its MFT saddle-point problem maps onto that of SSEP, with the corresponding on-shell actions equal and opposite. This establishes typicality without requiring an explicit solution of the saddle equations.


On the diffusive scale $\ell\equiv \sqrt{Dt}\gg1$, the SSEP
charge-transfer CGF is described by macroscopic fluctuation
theory (MFT)~\cite{Bertini_2015,Mallick_2022,PhysRevLett.133.117102,DerridaGerschenfeld2009MFT}. Introducing rescaled coordinates $x=j/\ell$
and $\tau=s/t$, the particle density $n(x,\tau)$ and
its conjugate field $\pi(x,\tau)$ evolve in the bulk
according to Hamilton's equations generated by
\begin{equation}
\mathcal H_{\rm SSEP}(n,\pi)
=-\partial_x n\,\partial_x\pi
+n(1-n)(\partial_x\pi)^2.
\label{eq:SSEPHamiltonian}
\end{equation}
The corresponding MFT action, including the initial
density fluctuations and the charge-transfer counting
field, is
\begin{align}
\mathcal S_{\rm SSEP}[n,\pi]&
=\int_0^1d\tau\int dx\,
[\pi\partial_\tau n-\mathcal H_{\rm SSEP}]\nonumber\\
&+\mathcal I_0[n_0]
-z\int_R dx\,[n(x,1)-n(x,0)],
\label{eq:SSEPaction}
\end{align}
where $n_0(x)=n(x,0)$ and $\mathcal I_0[n_0]$
is the large-deviation cost of the initial
coarse-grained density, determined by $\rho'_0$.
The leading scaled-CGF $\mu_{\rm SSEP}(z;\rho'_0)$ is
given by the negative of the action evaluated at its saddle.

The initial cost $\mathcal I_0[n_0]$ depends on
the statistics of the initial occupations, as well as their mean density. For example, the
half-filled equilibrium product state has
\begin{equation}
\mathcal I_0[n]=\int dx\,
\left[n\log(2n)
+(1-n)\log(2(1-n))\right].
\end{equation}
In contrast, a N\'eel state has the same
coarse-grained density $n_0(x)=1/2$, but
without initial density fluctuations:
$\mathcal I_0=0$ for this profile and
$+\infty$ otherwise. More generally,
$\mathcal I_0$ encodes the coarse-grained
density fluctuations of arbitrary
occupation-product initial states.

Although these saddle equations are not generally
solvable in closed form, the MFT boundary-value problem
is fully specified. We next show that the
large-time asymptotics of $\mathbb E_W Z_W^{-1}$
are governed by an equivalent saddle problem,
with the opposite on-shell action.

\textbf{Inverse moment and the mapping to SSEP}---Equation~\eqref{eq:gaussianDeterminants} expresses
subsystem-number statistics as a determinant,
whereas $Z_W$ generates charge transfer statistics.
To bring $Z_W$ into this form, we absorb the initial
counting factor $e^{-zN_R}$ into the initial
state, writing $e^{-zN_R}\rho'_0=A_0\widetilde\rho'_0$.
The normalization $A_0=\prod_j a_j$ and the
modified filling probabilities are
\begin{equation}
a_j=1-\varrho'_j+
\varrho'_j e^{-z\mathbf 1_{\{j\in R\}}},\quad 
\widetilde\varrho'_j =
\varrho'_j e^{-z\mathbf 1_{\{j\in R\}}}/a_j.
\label{eq:filteredinitial}
\end{equation}
Applying the Gaussian counting formula
Eq.~\eqref{eq:gaussianDeterminants} gives
\begin{equation}
Z_W=A_0\det\nolimits_R
[1+(e^z-1)\widetilde C_{R,W}(t)],
\label{eq:transferdeterminant}
\end{equation}
where $\widetilde C_W(t)=
U_W(t)\widetilde C_0U_W(t)^\dagger$
and $\widetilde C_0=
\operatorname{diag}(\widetilde\varrho'_j)$.

To represent the inverse determinant, introduce
independent complex Gaussian variables $\xi_j$ with
$\mathbb E_\xi|\xi_j|^2=\widetilde\varrho'_j$.
The Gaussian integral
$\mathbb E_\xi \exp[-\xi^\dagger M\xi]
=\det[1+\widetilde C_0 M]^{-1}$
allows us to rewrite Eq.~\eqref{eq:transferdeterminant}
in terms of the amplitudes
$\psi(t)=U_W(t)\xi$:
\begin{equation}
Z_W^{-1}
=A_0^{-1}\mathbb E_{\xi}
\exp\bigg[-(e^z-1)
\sum_{j\in R}|\psi_j(t)|^2\bigg].
\label{eq:inverseGaussian}
\end{equation}
This identity holds for each Brownian realization.

Remarkably, the auxiliary ``energies''
$\epsilon_j=|\psi_j|^2$ evolve autonomously.
Expanding the single-particle unitary to
second order in the noise, the contribution of a
single bond $(j,j+1)$ gives
\begin{equation}
\begin{aligned}
d\epsilon_j
&=D(\epsilon_{j+1}-\epsilon_j)\,dt+dM_j,\\
d\epsilon_{j+1}&=-d\epsilon_j,\qquad
(dM_j)^2=2D\epsilon_j\epsilon_{j+1}\,dt,
\end{aligned}
\end{equation}
where
$dM_j=-i\sqrt D
(\psi_j^*\psi_{j+1}\,d\overline W_j-\mathrm{c.c.})$.
The amplitude phases can be absorbed into
independent real Wiener processes, giving exactly
the classical Brownian energy process
(BEP)~\cite{GiardinaEtAl2009} for any fixed
initial $\xi$ (see End Matter for details).

Crucially, although the effective Wiener processes
depend on $\xi$, their probability law does not.
Thus, after averaging over $W$, the
energies $\epsilon_j$ have precisely the BEP law, with initial
energies $\epsilon_j(0)=|\xi_j|^2$ independently
exponentially distributed with means
$\widetilde\varrho'_j$. We therefore obtain
the exact classical representation
\begin{equation}
\mathbb E_W Z_W^{-1}
=A_0^{-1}\left\langle
e^{-(e^z-1)E_R(t)}
\right\rangle_{\rm BEP},
\label{eq:BEPidentity}
\end{equation}
where $E_R(t)=\sum_{j\in R}\epsilon_j(t)$,
and the BEP expectation includes the specified
initial exponential distribution.

We now formulate the MFT problem corresponding
to this classical process. On diffusive scales,
the coarse-grained energy density
$\varepsilon$ has diffusivity $D$
and mobility $2D\varepsilon^2$~\cite{Peletier2014}.
Its bulk action is $\mathcal S_{\rm BEP}^{\rm bulk}
=\int_0^1d\tau\int dx\,
[p\partial_\tau\varepsilon
-\mathcal H_{\rm BEP}]$, where $p$ is the response field and
\begin{equation}
\mathcal H_{\rm BEP}(\varepsilon,p)
=-\partial_x\varepsilon\,\partial_xp
+\varepsilon^2(\partial_xp)^2.
\label{eq:BEPHamiltonian}
\end{equation}

To incorporate the initial energy fluctuations,
we integrate the kinetic term by parts.
Its initial boundary contribution gives the factor
$\exp[\sum_j p_j(0)\epsilon_j(0)]$
in the path-integral weight.
Averaging over the independently
exponentially distributed initial energies yields
\begin{equation}
e^{\sqrt{Dt} B_0[p_0]}
=A_0^{-1}\prod_j
\left\langle e^{p_j(0)\epsilon_j(0)}
\right\rangle=\prod_j
\frac{1}{a_j(1-\widetilde\varrho'_j p_j(0))},
\label{eq:BEPinitialfunctional}
\end{equation}
where $p_j(0)=p(j/\ell,0)$.
The functional $B_0$ includes both
the initial energy fluctuations and the
normalization $A_0^{-1}$.

Retaining also the final boundary term from integrating
the kinetic action by parts and the energy-counting field, we obtain the complete BEP action
\begin{equation}
\begin{aligned}
\mathcal S_{\rm BEP}
={}&\int_0^1d\tau\int dx\,
[-\varepsilon\partial_\tau p
-\mathcal H_{\rm BEP}] -B_0[p_0]\\
&+\int dx\,
[p(x,1)+(e^z-1)\mathbf 1_R(x)]
\varepsilon(x,1),
\end{aligned}
\label{eq:BEPaction}
\end{equation}
where $\mathbf 1_R(x)$ denotes the indicator on $R$.

The key observation is that the transformation
\begin{equation}
n=\varepsilon(1-p),\qquad
\pi=\log(1-p),
\label{eq:BEPmap}
\end{equation}
previously identified for the Kipnis–Marchioro–Presutti (KMP) model and SSEP
\cite{PhysRevE.101.022209}, maps the BEP Hamiltonian to
$\mathcal H_{\rm BEP}=-\mathcal H_{\rm SSEP}$,
while the kinetic term transforms as
$-\varepsilon\partial_\tau p=n\partial_\tau\pi$.
Thus the bulk contributions to the two actions
differ by a sign.  Although such transformations generally complicate
the temporal boundary conditions
\cite{PhysRevLett.128.130602}, remarkably, in the
present problem the initial boundary functionals
also map onto one another.

To see this, integrating the kinetic term in Eq.~\eqref{eq:SSEPaction}
by parts gives the initial contribution
$\mathcal I_0[n_0]
-\int dx\,(\pi_0-z\mathbf1_R)n_0$.
Since $\mathcal I_0$ is the large-deviation
cost of the independent Bernoulli occupations,
minimizing over the initial density at fixed
$\pi_0$ gives, by Legendre duality,
\begin{equation}
\begin{aligned}
&\inf_{n_0}\left\{
\mathcal I_0[n_0]
-\int dx\,(\pi_0-z\mathbf1_R)n_0
\right\}\\
&\ \ =-\frac{1}{\sqrt{Dt}}
\sum_j\log\!\left(
1-\varrho'_j+
\varrho'_j e^{\pi_j(0)-z\mathbf1_{\{j\in R\}}}
\right)
=B_0[p_0].
\end{aligned}
\label{eq:BEPinitialcostmap}
\end{equation}
The last equality follows from
Eq.~\eqref{eq:BEPinitialfunctional},
using $1-p=e^\pi$ and Eq.~\eqref{eq:filteredinitial}, which together give $a_j(1-\widetilde\varrho'_j p_j)
=1-\varrho'_j+
\varrho'_j e^{\pi_j-z\mathbf1_{\{j\in R\}}}$.
Thus the initial SSEP boundary functional
reproduces that of the BEP, with the opposite
sign in the action.

The final boundary conditions likewise map onto one another, as shown in the End Matter. The corresponding BEP and SSEP saddle-point actions are therefore opposite. At zero counting field, both generating functions are governed by the usual hydrodynamic saddle. We make the standard MFT assumption that, at finite counting fields, the dominant saddles are smoothly connected to these zero-tilt solutions, without a dynamical phase transition to a competing saddle~\cite{Bertini_2015}. Together with Eq.~\eqref{eq:forwardSSEP}, this establishes Eq.~\eqref{eq:momentsgoal} and hence typicality of the leading charge-transfer CGF, and via Eqs.~\eqref{eq:entropyFCS} and \eqref{eq:vonNeumannFCS}, the diffusive growth of entanglement entropies.

\noindent\textbf{Discussion}---By introducing a new mapping of the noise-averaged inverse charge-transfer generating function to a classical Brownian energy process, whose hydrodynamic saddle-point problem maps onto that of SSEP, we have established the almost-sure typicality of charge-transfer statistics in individual noise realizations of QSSEP. This determines the long-time asymptotics of transient, diffusively growing charge-transfer fluctuations, both in equilibrium and following quantum quenches, in terms of classical SSEP transport statistics. Through a free-fermion identity relating entanglement to subsystem-number fluctuations, we further establish that the covariance-matrix eigenvalue distribution becomes self-averaging, and hence that the Rényi and von Neumann entropies are also typical. We have used these results to obtain asymptotically exact predictions for entanglement growth from domain-wall and Néel initial states, in excellent agreement with numerical simulations.

There are several directions for future work.
One is to extend these results to more general initial conditions, including non-diagonal product states and correlated or non-Gaussian states. Another is to apply the forward--inverse moment strategy to establish the typicality of transport fluctuations and entanglement dynamics in different geometries, including higher dimensions.

\noindent\textbf{Acknowledgments}---E.M. thanks Stefano Scopa, Adam Nahum, and Sarang Gopalakrishnan for helpful earlier discussions on related questions. E.M. acknowledges assistance from OpenAI's GPT-6 in strengthening the convergence in probability implied by Markov's inequality to almost-sure convergence. E.M. acknowledges support from the European Union (ERC, STAQQ, 101171399). 

Views and opinions expressed are however those of the authors only and do not necessarily reflect those of the European Union or the European Research Council Executive Agency. Neither the European Union nor the granting authority can be held responsible for them.

\makeatletter
\let\origbibsection\bibsection
\renewcommand{\bibsection}{%
  \begingroup
    \let\addcontentsline\@gobblethree
    \origbibsection
  \endgroup
}
\makeatother

\bibliography{Refs}

\makeatletter
\let\origbibsection\bibsection
\renewcommand{\bibsection}{%
  \begingroup
    \let\addcontentsline\@gobblethree
    \origbibsection
  \endgroup
}
\makeatother

\clearpage

\twocolumngrid
\begin{center}
    \textbf{\large End Matter}
\end{center}

\textbf{Entropy spectral identities}---For a Gaussian covariance eigenvalue $\nu$, the binary
entropy $h(\nu)=-\nu\log\nu-(1-\nu)\log(1-\nu)$
satisfies
\begin{equation}
h(\nu)=\frac14\int_0^\infty
\frac{\dd u}{\sinh^2(u/2)}
\log[1+4\nu(1-\nu)\sinh^2(u/2)].
\label{eq:EMbinaryEntropy}
\end{equation}
Differentiating the right-hand side with respect to
$\nu$ gives $\log[(1-\nu)/\nu]$, and both sides
vanish at $\nu=0$. Since the Gaussian determinant
implies
$\chi_W(u)+\chi_W(-u)
=\sum_a\log[1+4\nu_a(1-\nu_a)\sinh^2(u/2)]$,
summing over eigenvalues and subtracting the
initial value establishes Eq.~\eqref{eq:vonNeumannFCS}.

\textbf{Analytic continuation}---We have established typicality of the rescaled charge-transfer CGF for real counting fields. To obtain the leading integer R\'enyi entropies through Eq.~\eqref{eq:entropyFCS}, we must extend this result to the imaginary counting fields appearing in that identity. We first show that, for pure occupation-product initial states, the rescaled subsystem-number CGF converges almost surely to the analytic continuation of the classical SSEP current CGF. This follows from positivity of the charge-transfer probabilities and a standard Vitali--Hurwitz argument~\cite{Simon2022}. Since $N_R(0)$ is fixed for these states, $\Delta\chi_W(z,t)=\kappa_W(z,t;\rho_0)$. For each noise realization, the charge-transfer generating function admits the positive-probability representation
\begin{equation}
    Z_W(z,t)=\sum_q P_W(q,t)e^{zq},\quad P_W(q,t)\geq0.
    \label{eq:positive-prob}
\end{equation}
Consequently, $|Z_W(x+iy,t)|\leq Z_W(x,t).$

Moreover, the counting determinant in Eq.~\eqref{eq:gaussianDeterminants} is a product of factors $1-\nu_a+\nu_a e^z$, with $0\leq\nu_a\leq1$, whose zeros lie outside the strip $\mathcal D=\{z:|\operatorname{Im}z|<\pi\}$~\cite{PhysRevB.79.205315}. Thus $F_{t,W}(z)=(Dt)^{-1/2}\log Z_W(z,t)$ is analytic in $\mathcal D$.

Positivity and convexity of the real-field
CGF give,
\begin{equation}
|e^{F_{t,W}(x+iy)}|
\leq e^{F_{t,W}(x)}
\leq e^{\max\{F_{t,W}(-a),F_{t,W}(a)\}},
\end{equation}
for $|x|\leq a$ for any $a>0$. As shown in the Supplemental Material,
the forward and inverse moment asymptotics,
together with a bound on the local stochastic
evolution, establish almost-sure convergence
of $F_{t,W}(x)$ to the SSEP scaled current CGF for every finite real $x$.

For almost every noise realization, the bound
above therefore makes $e^{F_{t,W}}$
locally uniformly bounded throughout
$\mathcal D$. Vitali's theorem extends its
real-field convergence to locally uniform
convergence throughout the strip.
Since these functions are zero-free and
equal unity at $z=0$, Hurwitz's theorem
ensures that their limit is also zero-free,
allowing us to recover its analytic logarithm.
Thus $F_{t,W}$ converges almost surely
to the unique analytic continuation of the
SSEP CGF, establishing typicality of the
integer R\'enyi entropies.

The extension to mixed occupation-product
initial states is established in the
Supplemental Material using a general
bound on single-particle transmission
across the subsystem boundary.

\textbf{BEP dynamics}---We show that the ``energies''
$\epsilon_j=|\psi_j|^2$, where $\psi=U_W\xi$, evolve
autonomously as an exact classical stochastic process.
The contribution of a single bond $(j,j+1)$ to the
It\^o evolution of the amplitudes is
\begin{equation}
\begin{aligned}
d\psi_j
&=-i\sqrt{D}\,\psi_{j+1}\,d\overline W_j
  -\frac{D}{2}\psi_j\,dt,\\
d\psi_{j+1}
&=-i\sqrt{D}\,\psi_j\,dW_j
  -\frac{D}{2}\psi_{j+1}\,dt.
\end{aligned}
\end{equation}
Applying It\^o's rule gives
\begin{equation}
\begin{aligned}
d\epsilon_j\big|_{j,j+1}
&=-d\epsilon_{j+1}\big|_{j,j+1}
=D(\epsilon_{j+1}-\epsilon_j)\,dt+dM_j,\\
dM_j
&=-i\sqrt{D}\left(
\psi_j^*\psi_{j+1}\,d\overline W_j-\mathrm{c.c.}
\right).
\end{aligned}
\end{equation}
Remarkably, the amplitude phases can be absorbed
entirely into the Brownian noise. Writing
$\psi_j^*\psi_{j+1}
=\sqrt{\epsilon_j\epsilon_{j+1}}\,e^{i\theta_j}$,
we have $dM_j=\sqrt{2D\epsilon_j\epsilon_{j+1}}\,dB_j$,
where
\begin{equation}
dB_j=-\frac{i}{\sqrt{2}}\left(
e^{i\theta_j}d\overline W_j
-e^{-i\theta_j}dW_j\right).
\end{equation}
Although the angles $\theta_j$ depend on the
stochastic history, the $dW_j$ are fresh independent
increments. The $dB_j$ therefore define independent
real Wiener processes:
$dB_j\,dB_k=\delta_{jk}\,dt$, with zero conditional
mean. Thus, for any fixed initial $\xi$, the energies
evolve autonomously as the classical Brownian energy
process $\mathrm{BEP}(2)$~\cite{GiardinaEtAl2009},
without any approximation or averaging over
realizations.

For an ensemble of initial amplitudes $\xi$, however,
the effective Wiener processes $B^\xi$ depend on
$\xi$ through the phase rotations. Different initial
conditions evolved under the same $W$ therefore
experience different, generally correlated Wiener
processes. The Gaussian average over $\xi$ at fixed
$W$ cannot consequently be identified with an
average over initial conditions evolved under a
single BEP noise realization.

Crucially, this distinction disappears upon averaging
over $W$. For every fixed $\xi$, the rotated noise
$B^\xi$ has the same Wiener law, independent of
$\xi$. Thus, under the joint average over $W$ and
$\xi$, we may equivalently sample a single Wiener
process $B$ independently of $\xi$. Writing
$a=e^z-1$, Eq.~\eqref{eq:inverseGaussian} then gives
\begin{equation}
\begin{aligned}
\mathbb E_W Z_W^{-1}
&=A_0^{-1}\mathbb E_\xi\mathbb E_W
  e^{-aE_R(U_W\xi)}\\
&=A_0^{-1}\mathbb E_\xi\mathbb E_B
  e^{-aE_R^{\mathrm{BEP}}(t;|\xi|^2)}\\
&=A_0^{-1}\mathbb E_{\mathrm{BEP}}
  e^{-aE_R(t)}.
\end{aligned}
\end{equation}
Here the BEP starts from the independent exponential
energies $\epsilon_j(0)=|\xi_j|^2$. This establishes
the exact noise-averaged inverse-moment identity,
Eq.~\eqref{eq:BEPidentity}.

\textbf{Equivalence of the BEP and SSEP saddle problems}---We give further details of the mapping between the
BEP and SSEP boundary-value problems. In deriving
the BEP action, Eq.~\eqref{eq:BEPaction}, we integrated
the kinetic term by parts to incorporate the initial
energy distribution through the boundary functional
$B_0[p_0]$. To compare the two actions in the same
form, we also integrate the kinetic term of
the SSEP action, Eq.~\eqref{eq:SSEPaction}, by parts:
\begin{equation}
\begin{aligned}
\mathcal S_{\rm SSEP}
={}&\int_0^1d\tau\int dx\,
[-n\partial_\tau\pi-\mathcal H_{\rm SSEP}]
+\mathcal I_0[n_0]\\
&-\int dx\,(\pi_0-z\mathbf 1_R)n_0
+\int dx\,(\pi_1-z\mathbf 1_R)n_1.
\end{aligned}
\label{eq:EMSSEPparts}
\end{equation}

We first verify the bulk correspondence.
Under the transformation
$n=\varepsilon(1-p)$ and
$\pi=\log(1-p)$, Eq.~\eqref{eq:BEPmap},
we have $p=1-e^\pi$ and
$\varepsilon=ne^{-\pi}$. Substituting into
Eq.~\eqref{eq:BEPHamiltonian} gives
\begin{equation}
\begin{aligned}
\mathcal H_{\rm BEP}
&=(\partial_x n-n\partial_x\pi)
  \partial_x\pi+n^2(\partial_x\pi)^2\\
&=\partial_x n\,\partial_x\pi
-n(1-n)(\partial_x\pi)^2=-\mathcal H_{\rm SSEP}.
\end{aligned}
\end{equation}
The kinetic terms transform similarly: $-\varepsilon\partial_\tau p
=ne^{-\pi}e^\pi\partial_\tau\pi
=n\partial_\tau\pi$. Thus the bulk integrands of
Eqs.~\eqref{eq:BEPaction} and
\eqref{eq:EMSSEPparts} are precisely
opposites.

We must also compare the boundary terms.
Varying the final density $n_1$ in
Eq.~\eqref{eq:EMSSEPparts} gives
$\pi(x,1)=z\mathbf 1_R(x)$.
Similarly, varying the final energy density
$\varepsilon_1$ in Eq.~\eqref{eq:BEPaction}
gives $p(x,1)=-(e^z-1)\mathbf 1_R(x)$. Under the transformation
$\pi=\log(1-p)$, this becomes
$\pi(x,1)=z\mathbf 1_R(x)$, exactly
the SSEP final boundary condition.
Moreover, the final boundary terms
in both actions then vanish.

Finally, the initial boundary contributions
were shown in Eq.~\eqref{eq:BEPinitialcostmap}
to coincide under the same transformation,
with opposite signs in the two actions.
We have therefore established that the
complete BEP and SSEP actions are opposite
when evaluated at corresponding saddles, 
\begin{equation}
    \mathcal S_{\rm BEP}|_{\rm saddle} =-\mathcal S_{\rm SSEP}|_{\rm saddle}.
\end{equation}

The reversal of the bulk action under this field
transformation was previously discussed for KMP and
SSEP~\cite{PhysRevE.101.022209}. It does not imply a
reversal of the underlying large-deviation variational
principle: both processes have nonnegative current-history rate functionals, as is manifest in the current formulation of MFT. The response fields
arise from contour representations of the continuity
constraint, with the integration contours deformed
through the relevant saddles. 

\textbf{Covariance eigenvalue distribution}---The
subsystem number CGF determines not only the entanglement
entropies, but also the distribution of covariance
eigenvalues. To see this, write
$f_z(\nu)=\log[1+(e^z-1)\nu]$. Since $f_z(0)=0$, we have \(f_z(\nu_a)=\int_0^{\nu_a}d\nu\,f_z'(\nu)\). Equivalently,
\begin{equation}
f_z(\nu_a)=\int_0^1\dd\nu\,
\frac{e^z-1}{1+(e^z-1)\nu}
\,\mathbf 1_{\{\nu_a>\nu\}}.
\label{eq:f-integral-rep}
\end{equation}
Summing over the eigenvalues and subtracting their
initial contribution gives the exact identity
\begin{equation}
\frac{\Delta\chi_W(z,t)}{\sqrt{Dt}}
=\int_0^1\dd\nu\,
\frac{e^z-1}{1+(e^z-1)\nu}
\,\mathcal N_W(\nu,t),
\label{eq:EMspectralTransform}
\end{equation}
where $\mathcal N_W$ is the change in the cumulative
eigenvalue count between times $0$ and $t$, rescaled
by $\sqrt{Dt}$, as defined in Eq.~\eqref{eq:cumulativeSpectrum}.
This integral transform is invertible: expanding its
kernel around $z=0$ determines all moments of
$\mathcal N_W(\nu,t)$, which uniquely fix the
distribution. Since the rescaled subsystem-number
CGF converges almost surely to the corresponding
SSEP prediction, the change in the cumulative covariance eigenvalue count likewise converges to a nonrandom,
self-averaging limit determined by the SSEP CGF.

For the domain-wall initial state, the exact SSEP
generating function in Eq.~\eqref{eq:exactSSEPCGFs}
can be written as $\mu_{\rm DW}(z)=\frac1\pi
\int_{-\infty}^{\infty}\dd k\,f_z(e^{-k^2})$. Using the same integral representation Eq.~\eqref{eq:f-integral-rep} of $f_z$
and exchanging the order of integration gives
\begin{equation}
\begin{aligned}
\mu_{\rm DW}(z)
&=\int_0^1\dd\nu\, f_z'(\nu)
\left[
\frac1\pi\int_{-\infty}^{\infty}\dd k\,
\mathbf 1_{\{e^{-k^2}>\nu\}}
\right].
\end{aligned}
\end{equation}
Since $\Delta\chi_W(z,t)/\sqrt{Dt}\to\mu_{\rm DW}(z)$, comparing with Eq.~\eqref{eq:EMspectralTransform}
and using the uniqueness of the integral transform
identifies the limiting cumulative eigenvalue count:
\begin{equation}
\mathcal N_{\rm DW}(\nu)=\frac1\pi\int_{-\infty}^{\infty}\dd k\,
\mathbf 1_{\{e^{-k^2}>\nu\}}=\frac2\pi\sqrt{-\log\nu}.
\end{equation}
The corresponding domain-wall spectral moments were
previously obtained in Ref.~\cite{tonetto2026},
under an assumption of self-averaging.

For the N\'eel state, we again identify the cumulative eigenvalue distribution directly from the SSEP generating
function in Eq.~\eqref{eq:exactSSEPCGFs}. Writing $a=\sinh^2(z/2)$, changing variables to
$q=e^{-k^2}$, and integrating by parts gives
\begin{equation}
\mu_{\textrm{N\'eel}}(z)=\frac{\sqrt{2}}{\pi}\int_0^1\dd q\,
\sqrt{-\log q}\,\frac{a}{1+aq}.
\end{equation}
Now substitute $q=4\nu(1-\nu)$, with $0<\nu<1/2$.
Using
\begin{equation}
\frac{4a(1-2\nu)}{1+4a\nu(1-\nu)}
=f_z'(\nu)-f_z'(1-\nu),
\end{equation}
we obtain
\begin{align}
&\mu_{\textrm{N\'eel}}(z)\nonumber\\
&=\frac{\sqrt{2}}{\pi}
\int_0^{1/2}\dd\nu\,
\sqrt{-\log[4\nu(1-\nu)]}
\left[f_z'(\nu)-f_z'(1-\nu)\right].\nonumber\\
&=\frac{\sqrt{2}}{\pi}
\int_0^1\dd\nu\,f_z'(\nu)\,
\operatorname{sgn}\!\left(\frac12-\nu\right)
\sqrt{-\log[4\nu(1-\nu)]}\nonumber,
\end{align}
where to obtain the second equality we have changed variables $\nu\to1-\nu$ in the second term.

Comparing with Eq.~\eqref{eq:EMspectralTransform},
we identify the limiting cumulative eigenvalue
distribution,
\begin{equation}
\mathcal N_{\textrm{N\'eel}}(\nu)
=\frac{\sqrt2}{\pi}
\operatorname{sgn}\!\left(\frac12-\nu\right)
\sqrt{-\log[4\nu(1-\nu)]}.
\end{equation}
These are the limiting distributions quoted in Eq.~\eqref{eq:exactSpectra}, valid at fixed \(0<\nu<1\). Their divergence near the spectral endpoints reflects the accumulation of eigenvalues near their initial occupations.

\clearpage
\newpage

\widetext

\onecolumngrid

\input{SuppMat}

\end{document}

%% file: SuppMat.tex
\makeatletter

\begin{center}
\textbf{\large Supplemental Materials: Typicality of entanglement growth and transport fluctuations in quantum symmetric exclusion processes}

\vspace{3mm}

Ewan McCulloch

\vspace{2mm}

\textit{\small Laboratoire de Physique de l'École Normale Supérieure, CNRS,\\
ENS \& Université PSL; 24 rue Lhomond, 75005 Paris, France}

\makeatother


\makeatother

\end{center}

\begingroup
\makeatletter
\let\l@subsubsection\@gobbletwo
\tableofcontents
\makeatother
\endgroup

\makeatletter
\ltx@footnote@pop
\let\set@footnotewidth\set@footnotewidth@one
\let\compose@footnotes\compose@footnotes@one
\onecolumn@grid@setup
\makeatother

\setcounter{footnote}{0}

\setcounter{equation}{0}
\setcounter{figure}{0}
\setcounter{table}{0}
\setcounter{page}{1}
\makeatletter
\renewcommand{\theequation}{S\arabic{equation}}
\renewcommand{\thefigure}{S\arabic{figure}}

\section{Exact entanglement coefficients}
\label{sec:exact-entanglement-coefficients}

Here we derive the
entanglement coefficients quoted in
the main text using the
SSEP generating functions (Eq.~\eqref{eq:exactSSEPCGFs}) and the covariance
eigenvalue distributions derived in the End Matter.

For the domain-wall initial state, writing
$q=e^{-k^2}$ and substituting $\mu_{\rm DW}(z)$
into Eq.~\eqref{eq:entropyFCS} gives
\begin{equation}
\begin{aligned}
s_m^{\rm DW}
&=\frac{1}{\pi(1-m)}
\int_{-\infty}^\infty\dd k\,
\sum_{r=0}^{m-1}
\log[1-q+qe^{i\theta_{m,r}}]\\
&=\frac{1}{\pi(1-m)}
\int_{-\infty}^\infty\dd k\,
\log[(1-q)^m+q^m],
\end{aligned}
\label{eq:DWrenyiCoefficients}
\end{equation}
where we used
$\prod_r(1-q+qe^{i\theta_{m,r}})
=(1-q)^m+q^m$.

For the N\'eel state, we can obtain the entanglement
coefficients directly from the cumulative covariance
eigenvalue distribution derived above. Define the
binary R\'enyi entropy
\begin{equation}
h_m(\nu)=\frac{1}{1-m}
\log[\nu^m+(1-\nu)^m].
\end{equation}
The same integral identity used to relate the CGF
to the cumulative eigenvalue count gives
\begin{equation}
\frac{S_{m,W}(t)-S_{m,W}(0)}{\sqrt{Dt}}
=\int_0^1\dd\nu\,h_m'(\nu)\mathcal N_W(\nu,t).
\end{equation}
Using the N\'eel distribution in
Eq.~\eqref{eq:exactSpectra}, together with
$h_m'(1-\nu)=-h_m'(\nu)$, we obtain
\begin{equation}
s_m^{\textrm{N\'eel}}
=\frac{2\sqrt{2}}{\pi}
\int_0^{1/2}\dd\nu\,
\sqrt{-\log[4\nu(1-\nu)]}\,h_m'(\nu).
\label{eq:NeelRenyiCoefficients}
\end{equation}

The von Neumann coefficients follow by taking
$h_1(\nu)=-\nu\log\nu-(1-\nu)\log(1-\nu)$.
For the domain wall,
\begin{equation}
s_1^{\rm DW}
=\frac1\pi\int_{-\infty}^\infty
\dd k\,h_1(e^{-k^2}),
\end{equation}
while for the N\'eel state,
\begin{equation}
s_1^{\rm \textrm{N\'eel}}
=\frac{2\sqrt{2}}{\pi}
\int_0^{1/2}\dd\nu\,
\sqrt{-\log[4\nu(1-\nu)]}
\log\frac{1-\nu}{\nu}.
\end{equation}
Numerical evaluation gives the coefficients
reported in the main text.

\section{Convergence in probability of the charge-transfer cumulant generating function}
\label{sec:SMinprobability}

We show that the matching large-deviation exponents of the forward and inverse generating functions, established in the main text, imply convergence in probability of the rescaled charge-transfer cumulant generating function.

For any occupation-diagonal product initial state $\rho'_0$, fixed real counting field $z$, and $\ell=\sqrt{Dt}$, we have
\begin{equation}
\mathbb E_W Z_W^{\pm1}(z,t;\rho'_0)
=\exp\left[\pm\ell\mu_{\mathrm{SSEP}}(z;\rho'_0)+o(\ell)\right].
\end{equation}
Writing $\kappa_W=\log Z_W$ and $\mu=\mu_{\mathrm{SSEP}}(z;\rho'_0)$, Markov's inequality gives, for any $\varepsilon>0$,
\begin{align}
\mathbb P_W\left(\kappa_W>\ell(\mu+\varepsilon)\right)
&\le e^{-\ell(\mu+\varepsilon)}\mathbb E_W Z_W
=e^{-\varepsilon\ell+o(\ell)},\\
\mathbb P_W\left(\kappa_W<\ell(\mu-\varepsilon)\right)
&\le e^{\ell(\mu-\varepsilon)}\mathbb E_W Z_W^{-1}
=e^{-\varepsilon\ell+o(\ell)}.
\end{align}
Combining these inequalities,
\begin{equation}
\mathbb P_W\left(
\left|\frac{\kappa_W(z,t;\rho'_0)}{\sqrt{Dt}}
-\mu_{\mathrm{SSEP}}(z;\rho'_0)\right|>\varepsilon
\right)
\le 2e^{-\varepsilon\sqrt{Dt}+o(\sqrt{Dt})}
\longrightarrow 0.
\end{equation}
Thus the rescaled charge-transfer CGF converges in probability to the classical SSEP result. In fact, the probability of a finite deviation is stretched-exponentially suppressed in $\sqrt{t}$.

This establishes typicality at each sufficiently large time, but does not by itself guarantee convergence along an individual noise history for all sufficiently late times. In the following section, we exploit the locality of charge-transfer dynamics to strengthen this result to almost-sure convergence.

\section{Almost-sure typicality and analytic continuation}
\label{sec:SMalmostsure}

We explain why the real-field typicality established by the
forward and inverse generating-function calculations holds almost
surely, and why it extends to complex counting fields for arbitrary diagonal product states. There are two steps.
First, only hopping across the subsystem boundary changes the
charge-transfer generating function. This locality provides a
bound on its fluctuations over entire time intervals, upgrading
the large-time moment comparison to almost-sure typicality.
Second, we control the complex subsystem-number CGF using
single-particle transmission across the cut. 
The charge-transfer CGF of a domain wall initial
bounds this transmission for the \emph{same} Brownian
realization, allowing us to establish complex-field
typicality for both pure and mixed initial states.

We take the thermodynamic limit before the
long-time limit, working with a single cut on
an infinite chain. The stochastic bounds below
also hold at finite $L$, independently of system size.

\subsection{Almost-sure charge-transfer statistics at real counting fields}
\label{sec:SalmostsureCGF}

\subsubsection*{Why only the boundary bond matters}

Take any occupation-diagonal initial state $\rho'_0$ and a fixed
real counting field $z$. We wish to control
$Z_t\equiv Z_W(z,t;\rho'_0)$ not just at a specified time,
but throughout a microscopic time interval. Starting from the
definition in Eq.~\eqref{eq:transferFCS}, write
\begin{equation}
 Z_t=\Tr(A\tau_t),\qquad
 A=e^{zN_R},\quad
 \tau_t=\mathcal U_W(t)e^{-zN_R}\rho'_0\mathcal U_W(t)^\dagger.
 \label{eq:SMtiltedop}
\end{equation}
Since $[\rho'_0,N_R]=0$ and $z$ is real, $\tau_t$ is
positive. In particular, $Z_t>0$, and we can introduce the
normalized positive state
\begin{equation}
 \sigma_t=\frac{A^{1/2}\tau_t A^{1/2}}{Z_t}.
 \label{eq:SMnormalized}
\end{equation}
Its purpose is to express the stochastic change in $Z_t$
through bounded expectation values of operators near the cut.

To see this, write the hopping across the cut as
$T=c_1^\dagger c_0$, with site $1$ in $R$. The stochastic
Hamiltonian of Eq.~\eqref{eq:QSSEP} has increments
$\dd H_W=\sqrt D\sum_j(T_j\dd W_j+T_j^\dagger\dd\overline W_j)$.
The exact It\^o evolution of $\tau_t$ reads
\begin{equation}
\dd\tau_t={}-i[\dd H_W,\tau_t]
+D\sum_j\left(T_j\tau_tT_j^\dagger
+T_j^\dagger\tau_tT_j
-\tfrac12\{T_j^\dagger T_j+T_jT_j^\dagger,\tau_t\}\right)\dd t.
\label{eq:SMItoDensity}
\end{equation}
All bonds except the cut commute with $A$ and therefore
cancel on taking $\Tr(A\,\dd\tau_t)$. For the remaining bond,
$AT=e^zTA$ and $AT^\dagger=e^{-z}T^\dagger A$.

The resulting coefficients depend on just three local
expectation values,
\begin{equation}
 p_t=\Tr(T^\dagger T\sigma_t),\qquad
 q_t=\Tr(TT^\dagger\sigma_t),\qquad
 m_t=\Tr(T\sigma_t).
 \label{eq:SMpqm}
\end{equation}
For example, cyclicity and the first commutation relation give
$\Tr(AT\tau_tT^\dagger)=e^z\Tr(AT^\dagger T\tau_t)
=e^zZ_tp_t$; the corresponding anticommutator contributes
$-Z_tp_t$. Likewise, the commutator with the stochastic
hopping obeys
$\Tr(A[T,\tau_t])=2Z_t\sinh(z/2)m_t$.
Combining both directions of hopping, we obtain the exact
stochastic differential
\begin{equation}
\frac{\dd Z_t}{Z_t}
=D\big[(e^z-1)p_t+(e^{-z}-1)q_t\big]\dd t -2i\sqrt D\sinh(z/2)
\big(m_t\dd W_0-\overline m_t\dd\overline W_0\big).
\label{eq:SMItoZ}
\end{equation}

\subsubsection*{Bounding fluctuations over an entire time interval}

To exploit Eq.~\eqref{eq:SMItoZ}, write
$\dd Z_t/Z_t=a_t\dd t+\dd M_t$, with $\dd M_t$
the real, zero-mean noise increment. The It\^o multiplication
rule gives its quadratic variation,
\begin{equation}
 a_t=D[(e^z-1)p_t+(e^{-z}-1)q_t],\qquad
 (\dd M_t)^2=v_t\dd t,\qquad
 v_t=8D\sinh^2(z/2)|m_t|^2.
\label{eq:SMdriftvariance}
\end{equation}
The important point is that these quantities are uniformly
bounded. Indeed, $T^\dagger T=n_0(1-n_1)$ and
$TT^\dagger=n_1(1-n_0)$ project onto mutually exclusive
local configurations. Positivity of $\sigma_t$ therefore gives
\begin{equation}
 0\le p_t,q_t,\qquad p_t+q_t\le1,\qquad
 |m_t|^2\le p_tq_t\le\tfrac14.
 \label{eq:SMlocalbounds}
\end{equation}
In particular,
$v_t\le D(\cosh z-1)(p_t+q_t)$, so
\begin{equation}
 a_t\le D(e^{|z|}-1),\qquad -a_t+v_t\le D\sinh z\,(q_t-p_t)
 \le D\sinh|z|\le D(e^{|z|}-1).
\label{eq:SMbothdrifts}
\end{equation}
The combination $-a_t+v_t$ is relevant because It\^o's rule applied to the inverse gives
$\dd(Z_t^{-1})/Z_t^{-1}=(-a_t+v_t)\dd t-\dd M_t$.

Setting $c_z=D(e^{|z|}-1)$, Eq.~\eqref{eq:SMbothdrifts}
shows that multiplying $Z_t$ and $Z_t^{-1}$
by $e^{-c_zt}$ removes any possible positive
drift. Consequently, conditioned on the
Brownian noise history up to time $s$, the
expected values of these two rescaled
quantities at any later time $t\geq s$
cannot exceed their values at $s$.
Such positive stochastic processes are called
\emph{supermartingales}. Crucially, this
property allows us to bound fluctuations
over an entire time interval, rather than
only at a specified time.

We can now combine this property with the
self-averaging result. Namely, for an initial product state
$\rho'_0$, the forward and inverse moment results give,
for each fixed real $z$,
\begin{equation}
 \log\mathbb E_W Z_t^{\pm1}
 =\pm\ell\mu_{\rm SSEP}(z;\rho'_0)+o(\ell),
 \qquad\ell=\sqrt{Dt}.
 \label{eq:SMmoments}
\end{equation}
Let $\ell_n=\sqrt{Dn}$ and
$\mu=\mu_{\rm SSEP}(z;\rho'_0)$.
Doob's maximal inequality for nonnegative
supermartingales~\cite{Williams_1991}
provides a time-uniform analogue of Markov's
inequality: the probability that $Z_t$ exceeds
a threshold $A$ at any time in $[n,n+1]$
is bounded by $e^{c_z}\mathbb E_W Z_n/A$.
Applying the same inequality to $Z_t^{-1}$
bounds the probability that $Z_t$ becomes
unusually small during this interval. Together,
these yield
\begin{equation}
\begin{aligned}
\mathbb P_W\!\left(
 \sup_{n\le t\le n+1}|\log Z_t-\mu\ell_n|>
 \varepsilon\ell_n\right)
& \le e^{c_z-(\mu+\varepsilon)\ell_n}\mathbb E_W Z_n
 +e^{c_z+(\mu-\varepsilon)\ell_n}\mathbb E_W Z_n^{-1}\\
& \le 2\exp[-\varepsilon\ell_n+o(\ell_n)+c_z].
\end{aligned}
\label{eq:SMintervalbound}
\end{equation}

Because $\sum_{n\ge1}e^{-c\sqrt n}<\infty$
for any $c>0$, the Borel--Cantelli lemma
implies that, for any fixed $\varepsilon>0$,
deviations larger than $\varepsilon\ell_n$
occur on only finitely many intervals
$[n,n+1]$, almost surely.
In other words, for almost every noise
realization, there is a sufficiently late
time beyond which no further deviations
of this size occur.
Since this holds for arbitrarily small
$\varepsilon>0$, taking $\varepsilon\to0$
after $n\to\infty$ establishes~\footnote{Strictly, the
probability-one event obtained from the
Borel--Cantelli lemma may depend on
$\varepsilon$. We apply the argument to
the countable sequence $\varepsilon=1/k$,
$k\ge1$, and intersect the resulting
probability-one events.}
\begin{equation}
\lim_{n\to\infty}\sup_{n\le t\le n+1}
\frac{|\kappa_W(z,t;\rho'_0)-\mu\ell_n|}
{\ell_n}=0
\quad\text{almost surely}.
\end{equation}
Finally, $\ell_t/\ell_n=\sqrt{t/n}\to1$
uniformly for $n\le t\le n+1$, so
\begin{equation}
 \frac{\kappa_W(z,t;\rho'_0)}{\sqrt{Dt}}
 \longrightarrow\mu_{\rm SSEP}(z;\rho'_0)
 \quad\text{almost surely}.
 \label{eq:SMalmostsureReal}
\end{equation}
This holds for any fixed real $z$ and
any occupation-diagonal product initial state,
whether pure or mixed. Taking the intersection
of the probability-one events for rational $z$
gives simultaneous convergence on a dense
set of real counting fields.

\subsection{Analytic continuation of subsystem-number statistics}

We have established almost-sure convergence of the
rescaled charge-transfer CGF to the SSEP result for
real counting fields. In the End Matter, we extended
this result to complex fields for pure occupation-product
states. For these states, the initial subsystem particle
number is fixed, so the initial counting factor is
simply a scalar, and the change in subsystem-number
statistics coincides with charge-transfer statistics.
Positivity of the charge-transfer probabilities then
provides the bound needed for analytic continuation.

For general occupation-diagonal product states, the
initial subsystem particle number fluctuates, and this argument no longer applies. We instead bound
the change in the subsystem-number CGF directly,
using the total single-particle transmission across
the cut. This gives the required control at complex
counting fields for both pure and mixed initial states.

The Gaussian determinant already guarantees
that the change in subsystem number CGF $\Delta\chi_W(z,t)$ is analytic in
$\mathcal D=\{z:|\operatorname{Im}z|<\pi\}$,
since its factors $1-\nu+\nu e^z$,
$0\leq\nu\leq1$, have no zeros there.
To extend the real $z$ results to complex
$z$, it remains to show that this difference
is $O(\ell)$, uniformly on compact subsets
of $\mathcal D$, for almost every realization.
We establish this by bounding the change in
the CGF by the total single-particle transmission
across a cut.

\subsubsection*{Transmission bounds the change in the CGF}

Write the initial covariance as $C_0=C_L\oplus C_R$,
where $0\leq C_L,C_R\leq1$. Splitting the single-particle
propagator into left/right blocks 
\[
U_W=
\begin{pmatrix}
U_{LL}&U_{LR}\\
U_{RL}&U_{RR}
\end{pmatrix}.
\]
gives
\begin{equation}
 C_{R,W}(t)=U_{RL}C_LU_{RL}^\dagger
                 +U_{RR}C_RU_{RR}^\dagger.
 \label{eq:SMblockC}
\end{equation}
We wish to compare the final and initial covariance spectra
without obtaining a bound proportional to the subsystem
volume. The relevant quantity is the total single-particle
transmission from left to right,
\begin{equation}
 T_W=U_{RL}U_{RL}^\dagger,\qquad 0\leq T_W\leq1.
 \label{eq:SMtransmission}
\end{equation}

To isolate the effect of transmission, use the polar
decomposition $U_{RR}=VS$, where $V$ is unitary on $R$
and, by $U_W^\dagger U_W=1$,
\begin{equation}
 S=(U_{RR}^\dagger U_{RR})^{1/2}
  =(1-U_{LR}^\dagger U_{LR})^{1/2}.
\end{equation}
The unitary $V$ merely rotates the initial covariance
without changing its eigenvalues, while the deviation
of $S$ from the identity measures transmission out
of $R$. We can therefore compare $C_{R,W}(t)$ with
$VC_RV^\dagger$, which has exactly the initial
covariance spectrum. 

Substituting the polar decomposition of $U_{RR}$ yields $C_{R,W}(t)-VC_RV^\dagger
=U_{RL}C_LU_{RL}^\dagger
+V(SC_RS-C_R)V^\dagger$. Now apply the triangle inequality and unitary invariance of the trace norm:
\[
\begin{aligned}
\|C_{R,W}(t)-VC_RV^\dagger\|_1 &= \|U_{RL}C_LU_{RL}^\dagger
+V(SC_RS-C_R)V^\dagger\|_1\\
&\leq \|U_{RL}C_LU_{RL}^\dagger\|_1
+\|SC_RS-C_R\|_1\\
&\leq \operatorname{Tr}(U_{RL}U_{RL}^\dagger)
+\|SC_RS-C_R\|_1\\
&=\operatorname{Tr}T_W+\|SC_RS-C_R\|_1.
\end{aligned}
\]
The third inequality uses \(0\leq C_L\leq1\), which implies $0\leq U_{RL}C_LU_{RL}^\dagger
\leq U_{RL}U_{RL}^\dagger=T_W$.

To bound the remaining term, first write
\[
SC_RS-C_R=(S-1)C_RS+C_R(S-1).
\]
Using the triangle inequality and the trace-norm
bound $\|AB\|_1\leq\|A\|_1\|B\|_\infty$,
where $\|\cdot\|_\infty$ denotes the operator norm,
together with $0\leq C_R,S\leq1$, we obtain
\[
\begin{aligned}
\|SC_RS-C_R\|_1
&\leq\|(S-1)C_RS\|_1
+\|C_R(S-1)\|_1\\
&\leq\|S-1\|_1\|C_RS\|_\infty
+\|C_R\|_\infty\|S-1\|_1\\
&\leq2\|S-1\|_1.
\end{aligned}
\]

Furthermore, because \(0\leq S\leq1\), the matrix \(1-S\) is positive semidefinite, and therefore $\|S-1\|_1=\operatorname{Tr}(1-S)$. Hence
$\|SC_RS-C_R\|_1\leq2\operatorname{Tr}(1-S)$.

Finally, using $1-S\leq1-S^2=U_{LR}^\dagger U_{LR}$
and, by unitarity,
$\Tr U_{LR}^\dagger U_{LR}=\Tr T_W$, we have
\begin{equation}
\|C_{R,W}(t)-VC_RV^\dagger\|_1\leq3\Tr T_W.
\label{eq:SMpolarbound}
\end{equation}

To translate this spectral bound into one on the CGF,
define $f_z(s)=\log(1-s+se^z)$.
For any compact $K\subset\mathcal D$, its derivative
$f'_z(s)=(e^z-1)/(1-s+se^z)$ is uniformly bounded
for $z\in K$ and $s\in[0,1]$, since the denominator
has no zeros there. Denote this bound by $M_K<\infty$.
 
Since the CGF is the sum of $f_z$ over the
covariance eigenvalues, we can bound its change
by comparing the initial and final spectra.
For any two eigenvalues $s,s'\in[0,1]$,
\begin{equation}
 |f_z(s)-f_z(s')|
 =\left|\int_{s'}^s du\,f_z'(u)\right|
 \leq M_K|s-s'|.
\end{equation}
Applying this to the
initial and final eigenvalues, $\nu_a(0)$ and
$\nu_a(t)$ respectively, arranged in
increasing order, bounds the change in the CGF
by their total absolute difference. 
\begin{equation}
\sup_{z\in K}|\Delta\chi_W(z,t)|
\leq M_K\sum_a|\nu_a(t)-\nu_a(0)|
\label{eq:SMcomplex}
\end{equation}

We now use Mirsky's
inequality~\cite{Mirsky1957InequalitiesFN,Li1999}, $\sum_a|\lambda_a-\lambda_a'|\leq\operatorname{Tr}|X-X'|=\|X-X'\|_1$, for matrices $X$ and $X'$ with eigenvalues $\lambda_a$ and $\lambda_a'$ respectively, put in increasing order. Applying this in Eq.~\eqref{eq:SMcomplex} gives
\begin{equation}
\sup_{z\in K}|\Delta\chi_W(z,t)|
\leq M_K\|C_{R,W}(t)-VC_RV^\dagger\|_1 \leq3M_K\Tr T_W,
\label{eq:SMcomplexbound}
\end{equation}
where the last inequality follows from Eq.~\eqref{eq:SMpolarbound}

Thus the change in the CGF is controlled by
transmission across the cut, rather than by
the subsystem volume.

\subsubsection*{Domain-wall bound and analytic continuation}

It remains to show that $\Tr T_W$ grows at most on
the diffusive scale, for almost every realization.
We can obtain this directly from the real-field
typicality already established. Consider an
\emph{auxiliary} pure domain wall, with the left
half full and the right half empty, evolved using
the \emph{same} Brownian noise $W$. Its covariance
on $R$ is precisely $T_W$, and hence
\begin{equation}
 \kappa_W^{\rm DW}(1,t)
 =\Tr\log[1+(e-1)T_W]\geq\Tr T_W.
 \label{eq:SMdomainwallbound}
\end{equation}
The inequality follows from
$\log[1+(e-1)s]\geq s$ for $0\leq s\leq1$.

By Eq.~\eqref{eq:SMalmostsureReal}, applied to this
auxiliary initial state,
$\kappa_W^{\rm DW}(1,t)/\ell\to\mu_{\rm DW}(1)$
almost surely. Combining this with
Eq.~\eqref{eq:SMcomplexbound} gives, for every
compact $K\subset\mathcal D$,
\begin{equation}
 \sup_{z\in K}|F_{\ell,W}(z)|
 \leq3M_K\frac{\kappa_W^{\rm DW}(1,t)}{\ell}
 =O(1)\quad\text{almost surely},
 \label{eq:SManalyticlocalbound}
\end{equation}
where $F_{\ell,W}(z)=\ell^{-1}\Delta\chi_W(z,t)$.

We have now established that, for almost every
noise realization, $F_{\ell,W}$ is analytic and
locally bounded throughout $\mathcal D$.
Moreover, for each real $x$, the tilted-state
identity, Eq.~\eqref{eq:entropytransfer}, gives
$F_{\ell,W}(x)=\ell^{-1}\kappa_W(x,t;\rho_0(x))$,
whose almost-sure limit follows from
Eq.~\eqref{eq:SMalmostsureReal}.
Since the rational counting fields form a countable set,
convergence holds at all of them simultaneously for
almost every noise realization. This dense set is
sufficient to apply Vitali's theorem.

Vitali's theorem~\cite{Simon2022} then gives
locally uniform, almost-sure convergence
throughout $\mathcal D$:
\begin{equation}
 \frac{\Delta\chi_W(z,t)}{\sqrt{Dt}}
 \longrightarrow \mu_{\rm SSEP}(z;\rho_0(z)),
 \qquad |\operatorname{Im}z|<\pi.
 \label{eq:SMalmostsureComplex}
\end{equation}
Here the right-hand side is understood as the
analytic continuation of the real-field SSEP result.
This establishes the typicality of all fixed
integer R\'enyi entropies for both pure and mixed
occupation-product initial states.

\subsection{Von Neumann entropy: convergence of the counting-field integral}

The von Neumann entropy identity, Eq.~\eqref{eq:vonNeumannFCS},
involves an integral over all real counting fields.
Although the preceding results establish locally uniform
convergence of the rescaled CGF, we must also control
the contribution from arbitrarily large counting fields
before taking the long-time limit.

We obtain this control using the auxiliary domain wall
introduced above. Its charge-transfer CGF provides a
bound on the change in subsystem-number statistics
for any occupation-diagonal product initial state.

\subsubsection*{Bounding the CGF by an auxiliary domain wall}

For a fixed Brownian realization, write
$C_{R,W}(t)=U_{RR}C_RU_{RR}^\dagger+
U_{RL}C_LU_{RL}^\dagger$, as in Eq.~\eqref{eq:SMblockC}.
Recall that $T_W=U_{RL}U_{RL}^\dagger$ is the
restricted covariance of an auxiliary domain wall
evolved under the same noise realization.

We first show that, for $u\geq0$,
\begin{equation}
 \kappa_W^{\rm DW}(-u,t)
 \leq \Delta\chi_W(\pm u,t)
 \leq \kappa_W^{\rm DW}(u,t).
 \label{eq:SMentropyDWcomparison}
\end{equation}
To see this, begin with the expression for the evolved covariance, $C_{R,W}(t)
=U_{RR}C_RU_{RR}^\dagger
+U_{RL}C_LU_{RL}^\dagger$ together with \(0\leq C_L\leq1_L\). Here \(A\leq B\) again means that \(B-A\) is positive semidefinite. Multiplying the latter inequality on the left by \(U_{RL}\) and on the right by \(U_{RL}^\dagger\) preserves the ordering of Hermitian matrices: $0\leq U_{RL}C_LU_{RL}^\dagger
\leq U_{RL}U_{RL}^\dagger=T_W$. Adding \(U_{RR}C_RU_{RR}^\dagger\) throughout gives
\begin{equation}
 U_{RR}C_RU_{RR}^\dagger
 \leq C_{R,W}(t)
 \leq U_{RR}C_RU_{RR}^\dagger+T_W.
\end{equation}
The monotonicity and subadditivity of
$\Tr\log(1+bX)$ for positive matrices%
~\cite{Bhatia1997},
together with $U_{RR}^\dagger U_{RR}\leq1$,
give
\begin{equation}
 \Delta\chi_W(u,t)
 \leq\Tr\log[1+bT_W]
 =\kappa_W^{\rm DW}(u,t).
\end{equation}
Here $b=e^u-1\geq0$.

For the lower bound, let
$Q=1-U_{RR}^\dagger U_{RR}$ and
$\eta=1-e^{-u}=b/(1+b)$.
Using the determinant identity
$\det(1+AB)=\det(1+BA)$, we find
\begin{equation}
\begin{aligned}
 \frac{\det[1+bU_{RR}C_RU_{RR}^\dagger]}
        {\det[1+bC_R]} &=
 \det\left[
 1-Q^{1/2}\frac{bC_R}{1+bC_R}Q^{1/2}
 \right]\\
 &\geq\det(1-\eta Q)
 =\det(1-\eta T_W),
\end{aligned}
\end{equation}
where we used
$0\leq bC_R/(1+bC_R)\leq\eta$
and the fact that $Q$ and $T_W$ have
the same eigenvalues ($T_W=VQV^\dagger$ where $V$ is the unitary from the polar decomposition $U_{RR}=VS$). Consequently,
$\Delta\chi_W(u,t)\geq
\kappa_W^{\rm DW}(-u,t)$.

The same inequalities hold for $\Delta\chi_W(-u,t)$,
by applying the result to the hole covariance
$1-C_0$, using
$f_{-u}(s)=f_u(1-s)-u$.
Finally,
$-\kappa_W^{\rm DW}(-u,t)
\leq\kappa_W^{\rm DW}(u,t)$,
so Eq.~\eqref{eq:SMentropyDWcomparison} implies
\begin{equation}
 \left|\Delta\chi_W(u,t)+\Delta\chi_W(-u,t)\right|
 \leq2\kappa_W^{\rm DW}(u,t).
 \label{eq:SMentropyDWbound}
\end{equation}

\subsubsection*{A uniform diffusion bound}

We next establish a bound on the auxiliary
domain-wall CGF that holds even when the counting
field becomes large. By the exact one-replica
QSSEP--SSEP correspondence,
\begin{equation}
 \mathbb E_W e^{\kappa_W^{\rm DW}(u,t)}
 =\left\langle e^{uN_R(t)}
 \right\rangle_{\rm SSEP,DW}.
\end{equation}

We next use a known correlation inequality for SSEP
started from a deterministic configuration:
the probability of simultaneous occupation of
any set of sites is bounded by the product
of their individual occupation probabilities
\cite{BorceaBrandenLiggett2009},
\begin{equation}
 \left\langle\prod_{j\in A}n_j(t)\right\rangle
 \leq\prod_{j\in A}p_j(t).
\end{equation}
Since $e^{un_j}=1+(e^u-1)n_j$ for $n_j=0,1$,
expanding the generating function in products
of occupations, all with nonnegative coefficients
for $u\geq0$, gives
\begin{equation}
 \log\mathbb E_W e^{\kappa_W^{\rm DW}(u,t)}
 \leq\sum_{j\in R}
 \log[1+(e^u-1)p_j(t)],
 \label{eq:SMnegativeAssociation}
\end{equation}
where $p_j(t)$ is the mean SSEP occupation
at site $j$ for the domain-wall initial condition.

These mean occupations evolve by ordinary diffusion.
A random-walk exponential-moment bound gives
$p_j(t)\leq C_0e^{-cj/\ell}$ for $j\in R$,
where $\ell=\sqrt{Dt}\geq1$. For $u\geq0$, each term in the sum is at most $u$.
Moreover, using $\log(1+x)\leq x$, the contribution
from sites beyond a distance $j_*\sim\ell(1+u)$
is bounded by
\begin{equation}
 \sum_{j>j_*}\log[1+(e^u-1)p_j(t)]
 \leq C_0e^u\sum_{j>j_*}e^{-cj/\ell}
 =O(\ell).
\end{equation}
The remaining $O(\ell(1+u))$ sites each contribute
at most $u$. Consequently,
\begin{equation}
 \log\mathbb E_W e^{\kappa_W^{\rm DW}(u,t)}
 \leq C\ell(1+u)^2,\qquad u\geq0.
 \label{eq:SMDWmomentUniform}
\end{equation}
Importantly, this is a finite-time bound, independent
of any hydrodynamic approximation at large counting field.

We can upgrade it to an almost-sure bound using
the local stochastic evolution derived in Sec.~\ref{sec:SalmostsureCGF}.
For the auxiliary domain wall, the relative drift
of $Z_W^{\rm DW}(u,t)=e^{\kappa_W^{\rm DW}(u,t)}$
satisfies, from Eq.~\eqref{eq:SMbothdrifts},
\begin{equation}
 a_t=D[(e^u-1)p_t+(e^{-u}-1)q_t]\geq-D,
\end{equation}
independently of $u\geq0$.
Thus $e^{Dt}Z_W^{\rm DW}(u,t)$ is a positive
submartingale.

Let $\ell_n=\sqrt{Dn}$. Applying Doob's
submartingale maximal inequality~\cite{Williams_1991}
on $[n,n+1]$ and using
Eq.~\eqref{eq:SMDWmomentUniform},
we obtain, for $u\geq1$,
\begin{equation}
 \mathbb P_W\left(
 \sup_{n\leq t\leq n+1}
 \kappa_W^{\rm DW}(u,t)>
 K\ell_n(1+u)^2\right)\leq
 \exp\left[
 D-K\ell_n(1+u)^2
 +C\ell_{n+1}(1+u)^2
 \right].
\label{eq:SMDWmaximal}
\end{equation}
For sufficiently large fixed $K$, the right-hand side
is bounded by $e^{-c\sqrt n(1+u)^2}$ for some $c>0$.
These probabilities are summable jointly over integer
times $n$ and integer counting fields $u\geq1$.
The Borel--Cantelli lemma~\cite{Williams_1991} gives,
almost surely~\footnote{Strictly, Borel--Cantelli
applies to a countable collection of events.
We apply it at integer counting fields $u$,
then use the monotonicity of
$\kappa_W^{\rm DW}(u,t)$ to bound its value
between consecutive integers by its value
at the next integer. The resulting numerical
factor can be absorbed into $K$.},
\begin{equation}
 \kappa_W^{\rm DW}(u,t)
 \leq K\sqrt{Dt}(1+u)^2,
 \qquad u\geq1,
 \label{eq:SMDWuniformAlmostSure}
\end{equation}
uniformly for all sufficiently late times.

\subsubsection*{Convergence of the entropy integral}

Combining Eqs.~\eqref{eq:SMentropyDWbound}
and \eqref{eq:SMDWuniformAlmostSure},
we find, for every $\Lambda\geq1$,
\begin{equation}
 \limsup_{t\to\infty}\frac1{4\sqrt{Dt}}
 \int_\Lambda^\infty
 \frac{du}{\sinh^2(u/2)}
 \left|\Delta\chi_W(u,t)+\Delta\chi_W(-u,t)\right|\leq
 \frac K2\int_\Lambda^\infty
 \frac{(1+u)^2\,du}{\sinh^2(u/2)}
 \leq C'(1+\Lambda)^2e^{-\Lambda}
\label{eq:SMentropyTail}
\end{equation}
almost surely. The contribution from large
counting fields therefore vanishes uniformly
at late times as $\Lambda\to\infty$.

Near $u=0$, the numerator of the entropy integrand
vanishes quadratically. Indeed, local uniform
convergence of the rescaled analytic CGF also
gives uniform bounds on its derivatives, by
Cauchy's integral formula. Consequently,
$[\Delta\chi_W(u,t)+\Delta\chi_W(-u,t)]/\sqrt{Dt}
=O(u^2)$ uniformly for sufficiently late times.
This controls the apparent singularity of
$1/\sinh^2(u/2)$ at the origin.

We may therefore take the long-time limit
inside the entropy integral, first at fixed
$\Lambda$ and then sending $\Lambda\to\infty$.
Using the real-field limit established above,
we obtain
\begin{equation}
 \frac{\Delta S_{1,W}(t)}{\sqrt{Dt}}
 \longrightarrow
 \frac14\int_0^\infty\frac{du}{\sinh^2(u/2)}
 \left[
 \mu_{\rm SSEP}(u;\rho_0(u))
 +\mu_{\rm SSEP}(-u;\rho_0(-u))
 \right]
\end{equation}
almost surely. This establishes von Neumann
entropy typicality for both pure and mixed
occupation-diagonal product initial states.